 \documentclass[twocolumn,PRL,aps,psfig,showpacs,preprintnumbers,superscriptaddress]{revtex4}
\usepackage{graphicx}
\usepackage{epstopdf}
\usepackage{float}
\usepackage{color}
\usepackage{amsmath}

\begin{document}

\title{Magnetic fluctuations in the itinerant ferromagnet  LaCrGe$_{3}$ studied by $^{139}$La NMR }
\author{K. Rana}
\affiliation{Ames Laboratory, U.S. DOE, and Department of Physics and Astronomy, Iowa State University, Ames, Iowa 50011, USA}
\author{H. Kotegawa}
\affiliation{Graduate School of Science, Kobe University, Kobe 657-8501, Japan}
\author{R. R. Ullah}
\affiliation{Department of Physics, University of California, Davis, CA 95616, USA}
\author{J. S. Harvey}
\affiliation{Department of Physics, University of California, Davis, CA 95616, USA}
\author{S. L. Bud'ko}
\affiliation{Ames Laboratory, U.S. DOE, and Department of Physics and Astronomy, Iowa State University, Ames, Iowa 50011, USA}
\author{P. C. Canfield}
\affiliation{Ames Laboratory, U.S. DOE, and Department of Physics and Astronomy, Iowa State University, Ames, Iowa 50011, USA}
\author{H. Tou}
\affiliation{Graduate School of Science, Kobe University, Kobe 657-8501, Japan}
\author{V. Taufour}
\affiliation{Department of Physics, University of California, Davis, CA 95616, USA}
\author{Y. Furukawa}
\affiliation{Ames Laboratory, U.S. DOE, and Department of Physics and Astronomy, Iowa State University, Ames, Iowa 50011, USA}

\date{\today}

\begin{abstract} 

   LaCrGe$_3$ is an itinerant ferromagnet with a Curie temperature of $T_{\rm c}$ = 85 K  and exhibits an avoided ferromagnetic quantum critical point under pressure through a modulated antiferromagnetic phase as well as tri-critical wing structure in its temperature-pressure-magnetic field ($T$-$p$-$H$) phase diagram. 
     In order to understand the static and dynamical magnetic properties of LaCrGe$_3$, we carried out $^{139}$La nuclear magnetic resonance  (NMR) measurements.  
     Based on the analysis of NMR data, using the self-consistent-renomalization (SCR) theory, the spin fluctuations in the paramagnetic state are revealed to be isotropic ferromagnetic and three dimensional (3D)  in nature.
    Moreover, the system is found to follow the generalized Rhodes-Wohfarth relation which is expected in 3D itinerant ferromagnetic systems. 
    As compared to other similar itinerant ferromagnets, the Cr 3$d$ electrons and their spin fluctuations are characterized to have a relatively high degree of localization in real space.

\end{abstract}

\maketitle

 \section{Introduction} 
 
%

\begin{figure}[tb]
\includegraphics[width=\columnwidth]{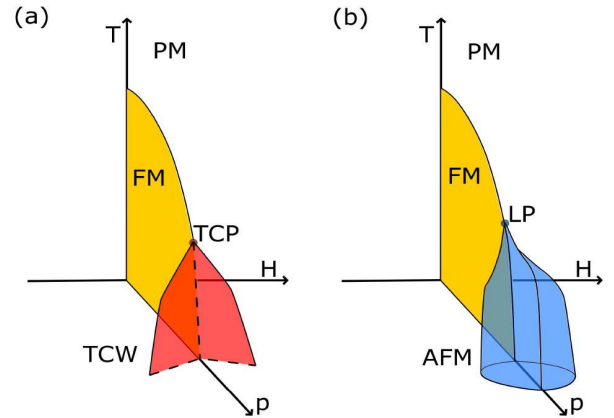} 
\caption{Generic temperature-pressure-magnetic field ($T$-$p$-$H$) phase diagrams for clean itinerant ferromagnets. (a) Schematic  $T$-$p$-$H$ phase diagram with a tricritical wing (TCW) structure. The second order phase transition (solid line) becomes first order (dashed line) at a tricritical point (TCP). Under magnetic field, the TCW structure shown in red emerge from the TCP.  (b) Schematic  $T$-$p$-$H$ phase diagram with an antiferromagnetic (AFM) phase in blue. The AFM phase  emerges from the Lifshitz point (LP) that is suppressed under $H$.} 
\label{fig:Fig0}
\end{figure}

    Recently much attention has been paid to itinerant ferromagnetic (FM) compounds because of the observations of unconventional superconductivity (SC) as well as  characteristic magnetic properties related to FM quantum criticality under application of pressure ($p$) and magnetic field ($H$)\cite{Canfield2016,Brando2016,K5,K6,K7,Aoki2001,Huy2007}.  
    Interestingly, in itinerant FM compounds, the FM quantum critical point (QCP) under $p$ is always avoided, which  has been of great interest in experimental and theoretical studies.
    Usually, when the second order paramagnetic (PM)-FM phase transition temperature ($T_{\rm c}$) is suppressed by the application of $p$, the order of the phase transition changes to the first order at the trictrtical point (TCP)  before $T_{\rm c}$  reaches 0 K at the quantum phase transition (QPT).
   This is known as the avoided QCP \cite{Taufour2010,Kotegawa2011,Kabeya2012}. 
   Here  the QCP is a second order quantum phase transition at $T$ = 0 K. \cite{Sachdev2011} 
 When the PM-FM transition becomes of the first order at the TCP in the $p$-$T$ plane, 
the application of magnetic field ($H$) leads to a tricritical wing (TCW) structure in the $T$-$p$-$H$ three dimensional phase diagram  [see, Fig. \ref{fig:Fig0}(a)] as found in UGe$_{\rm 2}$ \cite{Taufour2010,Kotegawa2011} and ZrZn$_{\rm 2}$\cite{Kabeya2012}. 
   A  PM-FM QCP  can also be avoided by the appearance  of an antiferromagnetic (AFM) ordered state under $p$ near the putative QCP, as actually observed in  CeRuPO\cite{Kotegawa2013,Lengyel2015} and MnP\cite{Cheng2015,Matsuda2016}.
    In this case, no wing structure has been reported and the AFM state is suppressed by the application of moderate $H$, as schematically shown in Fig. \ref{fig:Fig0}(b).

   Such interesting phase diagrams have been theoretically well studied, and PM-FM QCP in clean itinerant FM systems is suggested to be always avoided following the above two scenarios.  
     It has also been suggested that the coupling of quantum fluctuations of particle-hole excitations with the order parameter in metallic systems can generically render the prior second order phase transitions first order \cite{Belitz1999,Chubukov2004,Kirkpatrick2012,Kruger2014,Sang2014} or drive the system towards incommensurate ordering states \cite{Chubukov2004,Kirkpatrick2012}. 
   In addition, magnetic fluctuations may also play an important role to avoid FM QCP as some systems exhibit phases such as incommensurate magnetic ordered states \cite{Kotegawa2013,Lengyel2015,Cheng2015,Matsuda2016} and unconventional superconductivity \cite{K5,K6,K7,Aoki2001,Huy2007}.  
    Therefore,  the experimental characterization of the nature of quantum fluctuations, including magnetic fluctuations, in these classes of materials is important to illuminate the underlying mechanism that connects magnetism, quantum critically and superconductivity.

    Recently, the itinerant ferromagnet LaCrGe$_3$ has been discovered to be a new class of itinerant ferromagnets exhibiting the remarkable $T$-$p$-$H$ phase diagram where both the TCW structure  and AFM phase are observed \cite{Taufour2016,Kalu2017,Taufour2018}.            
    LaCrGe$_3$ crystallizes in the hexagonal BaNiO$_3$-type structure [space group $P6_3/mmc (194)$]  \cite{Bie2007}. 
   At ambient $p$, LaCrGe$_3$ is FM below the Curie temperature $T_{\rm c}$ = 85~K  with an ordered magnetic moment at low temperatures of 1.25 $\mu_{\rm B}$/Cr aligned along the $c$ axis \cite{Taufour2016}.
   This  small value of the magnetic moment compared with the effective moment above $T_{\rm c}$ ($\mu_{\rm eff}=$  2.4 $\mu_{\rm B}$/Cr) \cite{Lin2013} in the PM state suggests some degree of delocalization of the Cr 3$d$ spins.
   $T_{\rm c}$ can be suppressed with $p$ leading to a weakly modulated AFM state around 1.5 GPa and $T$ close to 50~K \cite{Taufour2016}.  
   Furthermore,  it was also reported to exhibit a TCP where the second order FM transition becomes of the first order at $T$ of around 40~K and $p$ close to 1.8 GPa \cite{Kalu2017} yielding a  TCW  structure under $H$.

    Motivated by the novel magnetic properties in LaCrGe$_3$, we carried out nuclear magnetic resonance (NMR)  measurement which is a powerful technique to investigate the magnetic and electronic properties of materials from a microscopic point of view.
    It is known that the temperature dependence of the nuclear spin-lattice relaxation rate (1/$T_1$) reflects the wave vector $q$-summed dynamical susceptibility. 
   On the other hand, NMR spectrum measurements, in particular the Knight shift $K$, give us information on local static magnetic susceptibility $\chi$. 
   Thus from the temperature dependence of 1/$T_1T$ and $K$, one can obtain valuable insights into spin fluctuations in materials. 
      In this paper, we report the results of $^{139}$La NMR measurements performed to investigate the spin fluctuations in LaCrGe$_3$.
     Our analysis, based on the self-consistent renomalization (SCR) theory, reveals that FM spin fluctuations due to Cr 3$d$ spins are of 3D nature. 
    In addition, LaCrGe$_3$ is well characterized by a relatively high degree of localization of 3$d$ spins although the system is itinerant. 
    Furthermore, the spin fluctuations are also revealed to have a more localized nature in real space  compared to other similar itinerant FM materials.

 \section{Experimental Details}
 
    Needle-like shaped single crystals of LaCrGe$_3$  were grown out of high temperature solutions, the details of which are reported in Ref. \cite{Lin2013}.
    Plural single crystals used for NMR measurements were placed in parallel on a glass plate (5$\times$5$\times$0.2 mm$^3$) to align their directions. 
   The crystalline $c$ axis and the $ab$ plane are parallel and perpendicular to the needle direction of the crystal, respectively. 
   An  NMR coil was tightly wound around the crystals including the glass plate to reduce a loss of the filling factor that is estimated to be about 0.6. 
    NMR  measurements of $^{139}$La ($I$ = $\frac{7}{2}$, $\frac{\gamma_{\rm N}}{2\pi}$ = 6.0146 MHz/T, $Q=$ 0.21 barns) nuclei were conducted using a lab-built phase-coherent spin-echo pulse spectrometer. 
        The $^{139}$La NMR spectra were obtained by sweeping $H$ at fixed frequencies  or by sweeping frequency under constant $H$. 
      $H$ was applied parallel to either the crystalline $c$ axis or the $ab$ plane.
    The zero-shift position corresponding to the Larmor field for each resonance frequency was determined by $^{31}$P NMR in H$_3$PO$_4$  solution or $^{63}$Cu NMR in Cu metal.

   The $^{139}$La nuclear spin-lattice relaxation rate (1/$T_{\rm 1}$) was measured with a saturation recovery method.
   $1/T_1$ at each temperature ($T$) was determined by fitting the nuclear magnetization $M$ versus time $t$  using the exponential function $1-M(t)/M(\infty) = 0.012e^{-t/T_1}+0.068e^{-6t/T_1}+0.206e^{-15t/T_1}+0.714e^{-28t/T_1}$,  where $M(t)$ and $M(\infty)$ are the nuclear magnetization at time $t$ after the saturation and the equilibrium nuclear magnetization at $t$ $\rightarrow$ $\infty$, respectively, for the case of magnetic relaxation \cite{Recovery}. 
    The observed recovery data in the paramagnetic state were well fitted by the function, indicating that the nuclear relaxation is mainly induced by fluctuations of the hyperfine field at the $^{139}$La site. 
    For the analysis of NMR data, we measured the magnetic susceptibility $\chi(T)$ of the single crystal at $H$ = 7 T applied parallel to the $c$ axis and to the $ab$ plane in a commercial Quantum Design superconducting quantum interference device magnetometer.

  \section{Results and discussion}
 \subsection{  $^{139}$La NMR spectrum}

\begin{figure}[tb]
\includegraphics[width=\columnwidth]{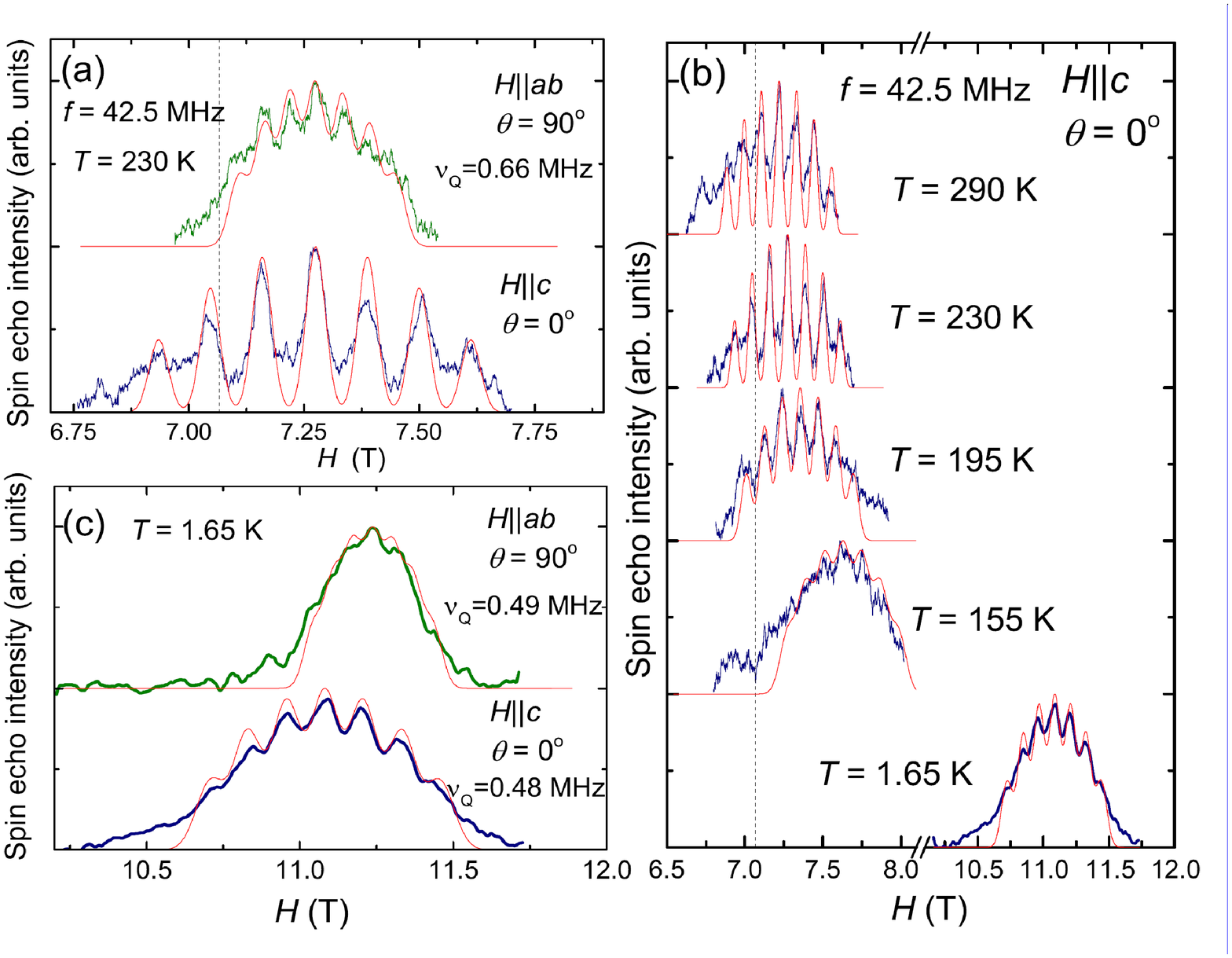} 
\caption{(a) Typical field-swept $^{139}$La-NMR spectra of LaCrGe$_3$ in the PM state  at $f$ = 42.5 MHz and $T$ = 230 K for $H||c$ and $H||ab$.
(b) Temperature dependence of $^{139}$La-NMR spectrum for $H||c$. 
(c) Field-swept $^{139}$La-NMR spectra in the FM state ($T$ = 1.65 K) for $H||c$ and $H||ab$.
Blue and green curves represent spectra for $H||c$ and $H||ab$, respectively, and  the red curves are the calculated spectra with $\nu_{\rm Q}$ = 0.66 MHz for the PM state and $\nu_{\rm Q}$ $\sim$  0.49 MHz for the FM state. 
The vertical dashed black lines in (a) and (b) represent the zero-shift position ($K$ = 0).
}
\label{fig:Fig1}
\end{figure}

      Figure\ \ref{fig:Fig1}(a) shows the field-swept $^{139}$La-NMR spectra of LaCrGe$_3$ at $T$ = 230 K for $H$ parallel to 
the $c$ axis ($H||c$) and to the $ab$ plane ($H||ab$). 
      The typical NMR spectrum for a nucleus with spin $I=7/2$ with Zeeman and quadrupolar interactions can be described by a nuclear spin Hamiltonian ${\cal{H}}=-\gamma\hbar\mathbf{I}\cdot\mathbf{H}_{\text{eff}}+\tfrac{h\nu_Q}{6}[3I_z^2-I^2+ \frac{1}{2}\eta(I_+^2 +I_-^2)]$,
where $\mathbf{H}_{\text{eff}}$ is the effective field at the nuclear site, $h$ is Planck's constant, and $\eta$ is the asymmetry parameter of  electric field gradient (EFG) at the nuclear site.  
   The nuclear  quadrupole frequency for $I=7/2$ nuclei is given by $\nu_{\rm Q} = e^2QV_{\rm ZZ}/14h$, where $Q$ is the nuclear quadrupole moment and $V_{\rm ZZ}$ is the EFG at the La site.
    When the Zeeman interaction is much greater than the Quadrupole interaction, this Hamiltonian produces a spectrum with a central transition line flanked by three satellite peaks on either side. 
   The observed spectra are well reproduced by simulated spectra (red lines) from the simple Hamiltonian with $\nu_{\rm Q}$ = 0.66~MHz and $\eta$ $\sim$ 0. 
    The tiny extra peaks around 6.80 T and 7.66 T for $H||c$ in Fig. 1(a) could be due to mis-orientation of some of the crystals while being attached on the glass plate and also may be due to slightly different qualities of the crystals. 
    The values of $\nu_{\rm Q}$ and $\eta$ estimated from the main seven peaks are found to be independent of temperature in the PM state above $T_{\rm C}$ = 85 K.
      From the spectrum analysis where  $\theta$s are found to be 0 and $\pi/$2 for $H \parallel c$ and $H \parallel ab$, respectively,  it is clear that the principal axis of the EFG at the La site is along the $c$ axis. 
    Since the La site in LaGeCr$_3$ does not have a local axial symmetry ($\bar6$$m$2), one may expect finite value of $\eta$. 
    However, $\eta$ is found to be very close to zero within our experimental uncertainty.

      As shown in Fig.~\ref{fig:Fig1}(b), with decreasing temperatures, each line becomes broader due to inhomogeneous magnetic broadening and the spectra show less clear features of the quadrupolar split lines below $\sim$ 155 K. 
     At the same time, nuclear spin-spin relaxation time $T_2$ becomes short at a wide range of temperatures close to $T_{\rm C}$. 
     Those make NMR spectrum measurements difficult below $\sim$ 110 K.  
     However,  when the temperature is decreased down to 1.65 K, well below $T_{\rm C}$ = 85 K, we were able to observe the $^{139}$La NMR spectrum in the FM state as shown at the bottom of Fig.~\ref{fig:Fig1}(b), where the spectrum largely shifted to higher magnetic field by $\sim$ 4 T. 
     The shift is due to the internal magnetic induction ($B_{\rm int}$) at the La site produced by the Cr spontaneous magnetic moments in the FM state.
     The $\nu_{\rm Q}$ $\sim$ 0.49 MHz estimated from the spectra under two different magnetic field directions ($H||c$ and $H||ab$) shown in Fig.~\ref{fig:Fig1}(c) is slightly smaller than 0.66 MHz observed in the PM state. 
      From the spectrum, the $B_{\rm int}$ is estimated to be $-$4 T and $-$4.2 T for $H||c$ and $H||ab$, respectively. 
      Here we took the zero-shift position ($K$ = 0) as the origin of $B_{\rm int}$.
     Unfortunately, the spectrum is measurable only around 1.6 K since its intensity decreases rapidly by raising the temperature.  

\begin{figure}[tb]
\includegraphics[width=\columnwidth]{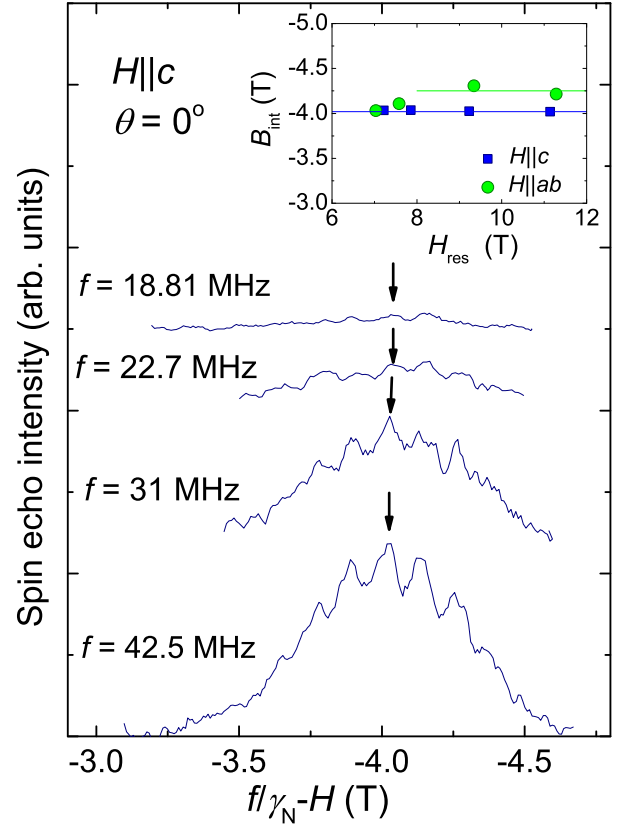} 
\caption{(a) $^{139}$La NMR spectra at various frequencies for $H||c$ at $T$ = 1.65 K. 
 Inset: Resonance magnetic field ($H_{\rm res}$) dependence of $B_{\rm int}$ for  $H||ab$~(green circles) and $H||c$~(blue squares). The arrows show the positions of the central transition line of the spectra measured at different frequencies}
\label{fig:Fig2}
\end{figure}

     In order to check whether the $B_{\rm int}$ is induced by the spontaneous Cr magnetic moments in the FM state and not due to NMR shift produced by the application of magnetic field, we determined the external magnetic field dependence of $B_{\rm int}$ by  measuring the spectra with different frequencies as shown in  Fig.~\ref{fig:Fig2}. 
     Although the signal intensity decreases with decreasing resonance frequency, we observed the spectrum down to $f$ = 18.81 MHz and found that $B_{\rm int}$ $\sim$ $-$4 T and $-$4.2 T for $H||c$ and $H||ab$ are nearly  independent of resonance frequency (i.e. external magnetic field),  confirming that these $B_{\rm int}$ values can be attributed to the hyperfine field at the La sites produced by the magnetic field independent spontaneous magnetic moment of the Cr ions in LaCrGe$_3$ in the FM state.

\begin{figure}[tb]
\includegraphics[width=\columnwidth]{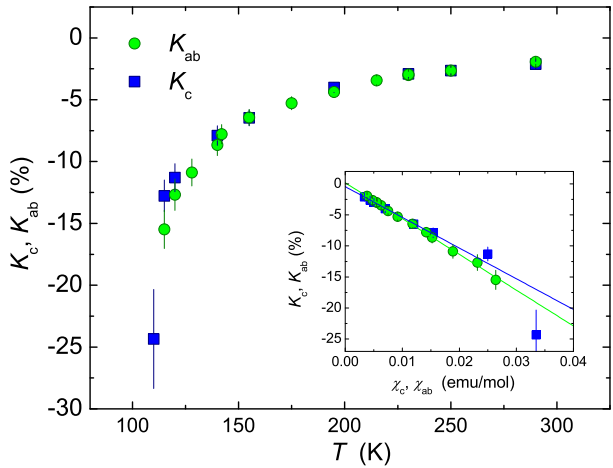} 
    \caption{Temperature dependence of  $^{139}$La Knight shift  for both $H||ab$ (green) and $H||c$ (blue) directions measured at $H$ $\sim$ 7 T. Inset: $K_{ab}$ vs. $\chi_{ab}$ and  $K_c$ vs. $\chi_c$ plots. The solid lines are linear fits as described in the text.  }
\label{fig:Fig3}
\end{figure}

   Figure~\ref{fig:Fig3} shows  the $T$ dependence of the $^{139}$La-NMR shift in the PM state for $H \parallel ab$ plane  ($K_{ab}$) and $H\parallel c$  axis ($K_{c}$)  determined from the simulated spectra, where both $K_{ab}$  and $K_{c}$ are nearly the same and decrease on lowering temperature.  
   The NMR shift consists of temperature  dependent spin shift $K_{\rm s}(T)$ and $T$ independent orbital shift $K_{\rm 0}$: $K(T)$ =$K_{\rm s}(T)$ + $K_{\rm 0}$ where $K_{\rm s}(T)$ is proportional to the spin part of magnetic susceptibility  $\chi_{\rm s}$($T$) via hyperfine coupling constant $A$, $K_{\rm s}(T)$  = $\frac{A\chi_{\rm s}(T)}{N_{\rm A}}$.  
      Here  $N_{\rm A}$ is Avogadro's number. 
    The hyperfine coupling constants are estimated to be $A_{ab}$ =  $-$32 $\pm$ 1  kOe/$\mu_{\rm B}$  and  $A_{c}$ = $-$27$\pm$ 1~kOe/$\mu_{\rm B}$ for $H||ab$ and $H||c$, respectively, from the slopes in the so-called $K$-$\chi$ plots shown in  the inset of  Figure~\ref{fig:Fig3}. 
    The intercepts for the fits are almost zero for both the directions. 
   This indicates that  the observed Knight shifts are mainly attributed to $K_{\rm s}(T)$.    
     $B_{\rm int}$ is proportional to $A_{\rm hf}$$<$$\mu$$>$ where $A_{\rm hf}$ is the  hyperfine coupling constant and $<$$\mu$$>$ is the ordered Cr magnetic moment.  
      Using $B_{\rm int}$ = $-$4.0 T and $-$4.2 T for $H||c$ and $H||ab$, respectively, and $A_{ab}$ =  $-$32 kOe/$\mu_{\rm B}$  and  $A_{c}$ = $-$27 kOe/$\mu_{\rm B}$, $<$$\mu$$>$ are estimated to be 1.30 $\mu_{\rm B}$  for $H||ab$ and 1.50 $\mu_{\rm B}$ for $H||c$ which are slightly higher but in good agreement with 1.22 $\mu_{\rm B}$ (along the $c$ axis) reported by the neutron diffraction measurements \cite{Cadogan2013} and 1.25 $\mu_{\rm B}$ (along the $c$ axis) from magnetization measurements \cite{Taufour2018,Lin2013}.

 \subsection{$^{139}$La spin lattice relaxation time} 

\begin{figure}[tb]
\includegraphics[width=\columnwidth]{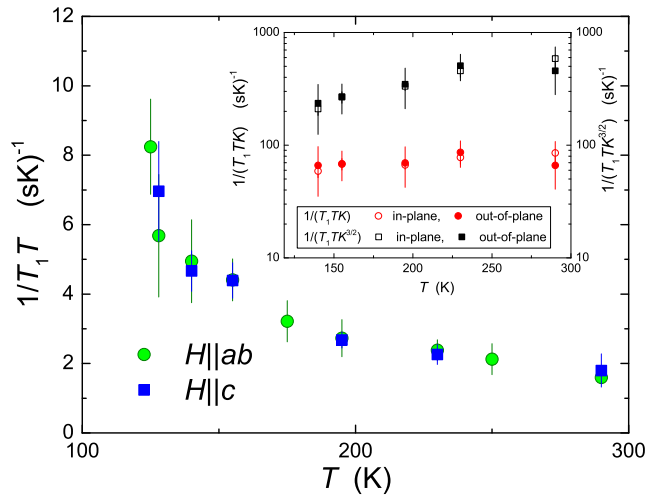} 
\caption{Temperature dependence of $^{139}$La 1/$T_1T$ for $H||ab$ (green circles) and $H||c$ (blue squares). 
    The inset shows the semi- log plots of 1/($T_1TK_{\rm s}$) and 1/($T_1TK^{3/2}_{\rm s}$)  for two different magnetic fluctuation directions (in-plane and out-of-plane directions).
 }
\label{fig:T1}
\end{figure}

      In order to investigate the magnetic fluctuations in LaCrGe$_3$, we  measured the $^{139}$La spin-lattice relaxation rate (1/$T_1$) at the peak position of the spectra for both the magnetic field directions. 
     Figure~\ref{fig:T1} shows the temperature dependence of 1/$T_1T$ where 1/$T_1T$ increases with decreasing temperature from room temperature to 125~K with no anisotropy in $T_1$.

    Based on the $T_1$ and the spin part of the Knight shift ($K_{\rm s}$) data,  we discuss the magnetic fluctuations in the PM state of LaCrGe$_3$. 
         First we tentatively employ the modified Korringa ratio analysis.
      In Fermi liquid picture, 1/$T_1T$ and $K_{\rm s}$ are determined by the density of states at the Fermi energy $\mathcal{D}(E_{\rm F})$. 
      The $T_1$ has relation with $K_{\rm s}$ that can be described as $T_1TK^2_{\rm s}$ = ($\hbar/4\pi k_{\rm B}$)($\gamma_{\rm e}/\gamma_{\rm n})^2$ $\equiv$ $S$.
       Here $\gamma_{\rm e}$ is the electronic gyromagnetic ratio.
      The Korringa ratio $\alpha$ ($\equiv$ $S/T_1TK_{\rm s}^2$) between an experimental value of $T_1TK^2_{\rm s}$ and the non-interacting electron system $S$ can reveal information about electron correlations in materials \cite{Moriya1963,Narath1968}.
  $\alpha\sim1$ represents the situation of uncorrelated electrons.  
     However, enhancement of  $\chi_{\rm s}$ ($\mathbf{q}\neq$ 0) increases 1/$T_1T$ but has little or no effect on $K_{\rm s}$ which probes only the uniform $\chi_{\rm s}$ ($\mathbf{q}$ = 0) where $\bf{q}$ represents  wave vector. 
       Thus $\alpha$ $>$ 1 indicates antiferromagnetic (AFM)  spin correlations. 
      In contrast, FM spin correlations produce $\alpha$ $<$ 1.
        Since 1/$T_1T$ probes magnetic fluctuations perpendicular to the magnetic field \cite{Moriya1963},  in general,  one should consider the Korringa ratio 1/($T_{1, \perp}TK^2_{{\rm s}, ab}$), where 1/($T_{1,\perp}T$) = 1/$(T_1T)_{H||c}$, when examining the character of magnetic fluctuations in the $ab$ plane (in-plane direction). 
     One also needs to consider the Korringa ratio 1/($T_{1, ||}TK^2_{{\rm s}, c}$) for magnetic fluctuations along the $c$ axis (out-of-plane direction). 
     Here, 1/($T_{1, \parallel}T$) is estimated from 2/$(T_{1}T)_{H||ab}$~$-$~$1/(T_{1}T)_{H||c}$. \cite{Wiecki2015}. 
     However, since both 1/$T_1$ and $^{139}K$ are nearly isotropic,  1/($T_{1, \perp}TK^2_{{\rm s}, ab}$) and  1/($T_{1, ||}TK^2_{{\rm s}, c}$) are almost the same, clearly indicating isotropic magnetic fluctuations in LaCrGe$_3$.
     $\alpha_\parallel$ and $\alpha_\perp$ decrease from $\sim$0.07 at room temperature to less than $\sim$0.008 around 120 K, indicating dominant FM spin correlations between Cr spins in the compound.

    It should be noted that, however, the Korringa analysis usually applies for PM materials where electron-electron interaction is weak.  
    Since  LaCrGe$_3$ exhibits a FM order,  we also analyze NMR data based on self-consistent renormalization (SCR) theory.
     As shown above, the magnetic fluctuations are governed by FM spin correlations. 
   In this case, according to SCR theory for weak itinerant ferromagnets,  1/($T_1TK_{\rm s}$) and 1/($T_1TK^{3/2}_{\rm s}$) are expected to be independent of $T$  for three dimensional (3D) or  two-dimensional (2D) FM spin fluctuations, respectively \cite{Moriya1974,Hatatani1995}. 
       The inset of Fig.~\ref{fig:T1} shows  the $T$ dependence of 1/($T_1TK_{\rm s})$ and 1/($T_1TK_{\rm s}^{3/2})$ for the two directions.
      Both the 1/($T_{1,\parallel}TK_{\rm s,c}$) and 1/($T_{1,\perp}TK_{{\rm s},ab}$)  are nearly constant, while both the 1/($T_{1,\parallel}TK_{{\rm s},c}^{3/2}$) and 1/($T_{1,\perp}TK_{{\rm s},ab}^{3/2}$) increase with increasing temperature. 
      This indicates that  the FM spin fluctuations are characterized as 3D  in nature.

 \subsection{Spin fluctuations }

    The Curie Weiss behavior of 1/$T_1T$ in itinerant ferromagents is well described in the SCR theory in terms of 3D FM spin fluctuations \cite{Moriya1973,Kontani1976,Takahashi1984,Takahashi1986,Yoshimura1987}, as described above. 
    According to SCR theory, there are two important parameters to characterize spin fluctuation for itinerant ferromagnets: $T_0$ and $T_{\rm A}$ corresponding to the widths of the spin fluctuations in frequency ($\omega$) space at wave vector $(q)$ = 0 and the width of the distribution of static susceptibility in $q$ space at $\omega$ = 0, respectively \cite{Takahashi2013}.
      The former can be obtained from the following equation of 1/$T_1T$  as derived by Corti \textit{et al.} \cite{Corti2007}, 
\begin{equation}
\frac{1}{T_1T}=\frac{3\hbar \gamma_{\rm N}^2 H_{hf}}{16 \pi \mu_{\rm B}}\frac{K}{T_0} 
\label{eq:SCR_T0}
\end{equation}
Here $H_{\rm hf}$ 
is the hyperfine field experienced at the La site per spin \cite{Corti2007}.
   From the experimentally determined values, $T_0$ is estimated to be 89 K and 75 K for the in-plane and out-of-plane directions of spin fluctuations, respectively. 
   On the other hand, $T_{\rm A}$ can be estimated using the following equation given by Takahashi \cite{Takahashi2013}:
\begin{equation}
T_A=20C_{4/3}\Big[\frac{T_c}{p_s^{3/2}T_0^{1/4}}\Big]^{4/3}
\label{eq:SCR_TA}
\end{equation}
where $T_{\rm c}$ = 85 K, $p_{\rm s}$ is the saturated moment   and C$_{4/3}$=1.006089 is a constant. 
Utilizing these values, $T_{\rm A}$ is estimated to be 998 K and 793 K for the in-plane and out-of-plane directions of spin fluctuations, respectively.  
   The estimated values of $T_{\rm A}$ seem to be comparable or slightly greater than those in uranium based compounds such as URhGe ($T_{\rm A}$ = 568  K) \cite{Tateiwa2017} and UGe$_2$ ($T_{\rm A}$ = 442 K) \cite{Tateiwa2017}, but are smaller than those in 3$d$ electron itinerant ferromagnets such as  ZrZn$_2$ ($T_{\rm A}$ = 7400 K) \cite{Takahashi2013}, NiAl$_3$ ($T_{\rm A}$ = 3670 K) \cite{Takahashi2013}  and MnSi ($T_{\rm A}$ = 1690  K) \cite{Corti2007}.

\begin{figure}[tb]
\includegraphics[width=\columnwidth]{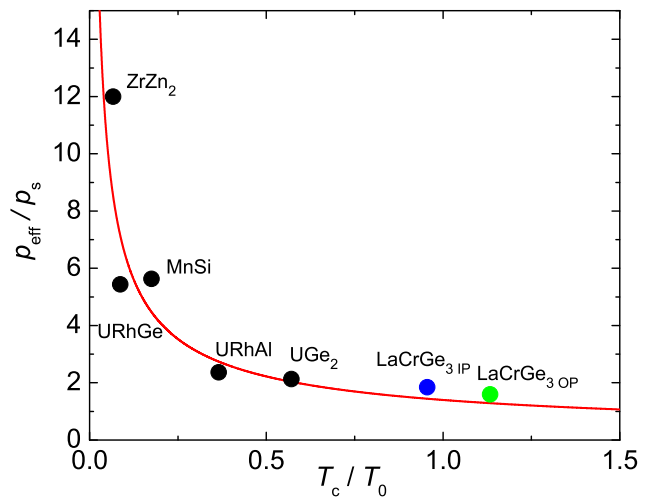} 
\caption{ Generalized Rhodes-Wohlfarth plot for the in-plane (blue) and out-of-plane (green) directions for LaCrGe$_3$  along with other itinerant ferromagnets that avoid FM-PM QCP through tricritical wings under pressure. 
     The values of parameters are estimated from Ref. \cite{Takahashi2013} for ZrZn$_2$ and  MnSi and Ref. \cite{Tateiwa2017} for UGe$_{\rm 2}$,  URhGe and URhAl. 
     The red  line represents the generalized Rhodes-Wohlfarth relation in the 3D system: $p_{\rm eff}/p_{\rm s}$ = 1.4 ($T_{\rm C}$/$T_0$)$^{-2/3}$ \cite{Takahashi2013}. 
}
\label{fig:SCR}
\end{figure}


     In order to compare  LaCrGe$_3$ with other clean itinerant ferromagnets that avoid the FM QCP with the TCW structure, we plotted their spin fluctuation parameters in Fig.~\ref{fig:SCR} along with those in  LaCrGe$_3$. 
   The plot is  known as  the Generalized Rhodes-Wohlfarth plot.
     The $x$ axis of this plot is the ratio of $T_{\rm C}$ and $T_0$ and the $y$ axis the ratio of the effective paramagnetic moment ($p_{\rm eff}$) and $p_{\rm s}$. 
     According to SCR theory, in the case of itinerant ferromagnets, magnetic fluctuations contribute to $p_{\rm eff}$ in PM state.
     Therefore, one can expect that the ratio of $p_{\rm eff}$/$p_{\rm s}$ becomes large, when spin fluctuations are important in these systems. 
      As for $T_0$, based on SCR theory, $T_{\rm C}$/$T_0$ is close to unity in the case of localized ferromagnets, while the ratio becomes less than unity if itinerant character becomes significant. 
       These situations are predicted theoretically through the generalized Rhodes-Wolfarth relation of  $p_{\rm eff}/p_{\rm s}$ = 1.4 ($T_{\rm C}/T_0$)$^{-2/3}$ (red line),  and the relation has shown great empirical agreement with materials having 3D FM fluctuations \cite{Takahashi2013}. 
       As shown in the  Fig.~\ref{fig:SCR},  LaCrGe$_3$ and other typical itinerant ferromagnets seem to follow the relation.
   However, the values of LaCrGe$_3$ for both the in-plane and out-of-plane directions of magnetic fluctuations are located close to unity, indicating   
a  relatively high degree of localization in Cr $3d$ electrons even though the system is itinerant.
    In addition, the results also indicate that the spin fluctuations in LaCrGe$_3$ show a more localized nature in real space than the other similar itinerant FM materials compared. 


   \section{Summary}

    In summary, we carried out $^{139}$La NMR in the itinerant ferromagnet LaCrGe$_3$  to  characterize the magnetic properties from a microscopic point of view.
   The  principal axis of the electric field gradient at the La site has been shown to be along the $c$ axis with $\nu_Q$ of 0.66 MHz in the PM state and about 0.49 MHz in the FM state. 
     $^{139}$La NMR spectra measurements in the FM state confirmed the FM ordered state below $T_{\rm C}$ = 85 K with a magnetic moment of $\sim$ 1.4
 $\mu_{\rm B}$/Cr, consistent with previous reports.
    Based on the Korringa ratio analysis and  the  SCR theory using the results of 1/$T_1T$ and $K$, spin fluctuations in the PM state are revealed to be isotropic FM and three dimensional in nature.
    In addition, the system is found to follow the generalized Rhodes-Wohfarth relation which is expected in 3D itinerant FM systems. 
   $T_{\rm C}$/$T_0$ is found to be close to unity, indicating that there is a relatively high degree of  localization in the 3$d$ Cr electrons even though the system is itinerant.  
   It would be interesting if these uniqueness in magnetic fluctuations in LaCrGe$_3$ is related to the appearance of both tricritcal wings and AFM state under pressure and magnetic fields. 
    Analysis of spin fluctuations in those materials where FM QCP is avoided by the appearance of AFM phase, and further studies on LaCrGe$_3$ under pressure are important to investigate this connection, which are now in progress.

 \section{ Acknowledgments} 

   The authors would like to thank Q.-P. Ding, R. Takeuchi and Y. Kuwata, for help in conducting experiments and T. Matsui for fruitful discussions.      
  The research was supported by the U.S. Department of Energy, Office of Basic Energy Sciences, Division of Materials Sciences and Engineering. 
    Ames Laboratory is operated for the U.S. Department of Energy by Iowa State University under Contract No.~DE-AC02-07CH11358.
      Part of the work was supported by the Japan Society for the Promotion of Science KAKENHI Grant Numbers JP15H05882, JP15H05885, JP15K21732, and  JP18H04321 (J-Physics).
     K. R. also thanks the KAKENHI: J-Physics for the financial support that provided an opportunity to be a visiting scholar at Kobe University.

\end{document}